\documentclass[pdflatex,sn-mathphys-num]{sn-jnl}

\usepackage{graphicx}%
\usepackage{multirow}%
\usepackage{amsmath,amssymb,amsfonts}%
\usepackage{amsthm}%
\usepackage{mathrsfs}%
\usepackage[title]{appendix}%
\usepackage{xcolor}%
\usepackage{textcomp}%
\usepackage{manyfoot}%
\usepackage{booktabs}%
\usepackage{algorithm}%
\usepackage{algorithmicx}%
\usepackage{algpseudocode}%
\usepackage{listings}%
\usepackage[T1]{fontenc}
\usepackage[utf8]{inputenc}

\begin{document}

\title[Article Title]{Correction of the influence of rolling shutter detectors on determining the velocity of meteors from video recordings}


\author*[1]{\fnm{Lukáš} \sur{Shrbený}}\email{shrbeny@asu.cas.cz}
\author[1]{\fnm{Jiří} \sur{Borovička}}
\author[1]{\fnm{Pavel} \sur{Spurný}}
\author[1]{\fnm{Jan} \sur{Mánek}}
\author[1]{\fnm{Jan} \sur{Fuchs}}

\affil[1]{\orgdiv{Interplanetary Matter Department}, \orgname{Astronomical Institute of the Czech Academy of Sciences}, \orgaddress{\street{Fričova 298}, \city{Ondřejov}, \postcode{25165}, \country{Czech Republic}}}


\abstract{Modern video cameras use either global or rolling shutter to create individual video frames. In case of global shutter, the whole frames are exposed and read out at the same time. In case of rolling shutter, the exposure and readout proceed sequentially line by line. Such cameras are now commonly used for meteor observations. It is then necessary to perform a correction for this type of image reading in order to correctly and accurately determine the velocity of recorded meteors. We provide here the correct procedure. The correction must include the value of the rolling period of the sensor, which is generally not provided in camera specifications. We discuss four methods of determining the rolling period experimentally. The most accurate method uses the New EXposure Timing Analyser (NEXTA) of \cite{kam23}. We list rolling periods measured for several cameras.}


%
%
%




\maketitle

\section{Introduction}\label{int}
Over the past few decades, there has been a huge increase in the use of digital video cameras that use Complementary Metal-Oxide-Semiconductor (CMOS) sensors with rolling shutters. Unlike reading the entire sensor area at the same time (global shutter detectors), they are based on sequential reading of individual lines of the image sensor. The advantage of this type of scanning is the possibility of higher sensor resolution and higher frame rate. The disadvantage is the distortion of fast-moving objects, especially those moving in the direction of or against the direction of sensor readout. Such objects include meteors and fireballs. Today, there are several meteor networks, such as Allsky7 \citep{han20} or Global Meteor Network (GMN) \citep{vid21}, that use video cameras with rolling shutter CMOS sensors as the sole recording device to record meteors and subsequently determine their atmospheric trajectory, velocity, radiant, and heliocentric orbit. Precise correction of the rolling shutter effect is therefore important for correctly determining the velocity of the meteor, which is an important parameter for determining the heliocentric orbit. The only paper describing the influence of rolling shutter cameras on determining the velocity of meteors was published by \cite{kuk18} and forms the basis for the correction procedure used by GMN and Allsky7 networks. Unfortunately, this paper assumes a simplifying condition, namely that the sensor readout time is equal to the inverse of the frame rate, which reduces the correction to the relative vertical position of the meteor on the sensor.

The Czech Fireball Network (CFN) which is a part of the European Fireball Network (EN)  uses security internet protocol (IP) video cameras as supplementary instrument besides photographic cameras. The main purpose of using these IP cameras during the night is to record the fragmentation of fireballs and to increase the accuracy of dynamics of distant fireballs \citep{shr26}. A huge advantage of the Czech part of EN is the fact that it uses photographic cameras that have been proven over decades to detect fireballs and provide accurate trajectories and velocities of them \citep[e.g.][]{bor22,spu24}. These cameras regularly interrupt the exposure using a liquid crystal (LCD) shutter so that the time stamps behave in the same way as on a video camera with a global shutter detector. This makes it possible to directly compare the dynamics from photographic and video cameras and thus verify the accuracy of the rolling shutter correction. The conclusion of this comparison was that the correction by \cite{kuk18} is not sufficient for all fireballs. In the following sections, we will describe our rolling shutter correction procedure, which includes the value of the sensor rolling period and also methods how to determine the rolling period.

\begin{figure}[h]
\centering
\includegraphics[width=0.99\textwidth]{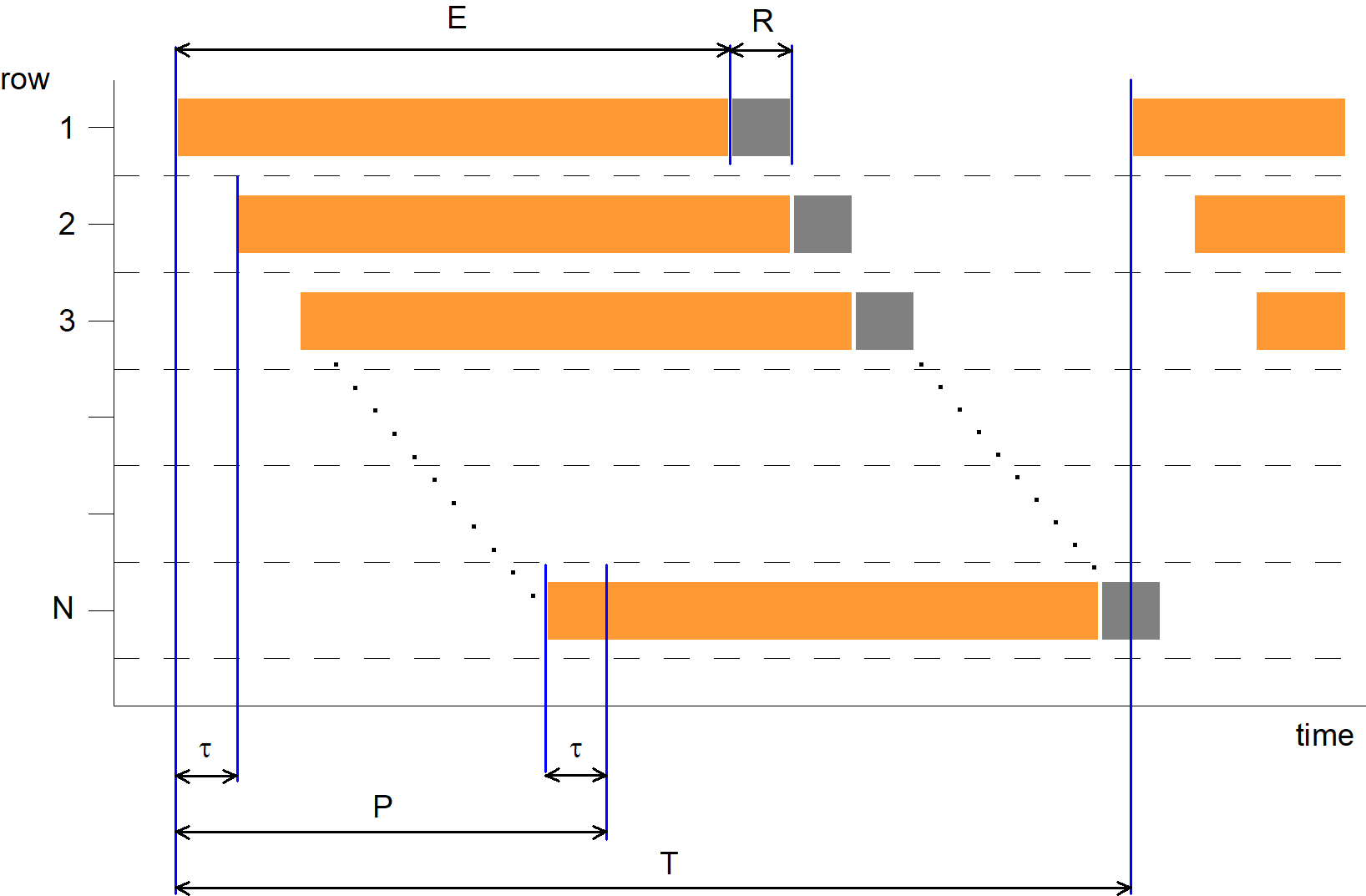}
\caption{Schematic timeline of the rolling shutter showing individual time periods. See the text for explanation.}\label{f_rs}
\end{figure}

\section{Definition of basic terms}\label{def}
We consider a CMOS sensor with $N \times M$ pixels, where $N$ is the number of rows and $M$ is the number of columns. In the following, the number of columns is not important. The video is produced with $F$ frames per second (FPS). The rolling is assumed to proceed along rows from top to the bottom, i.e. the exposure of a frame starts in the first row and proceed down. The coordinate system $(x,y)$ has the origin at the upper left corner. The $x$ coordinate is measured horizontally from left to right and the $y$ coordinate is measured vertically from top to bottom. The coordinates are measured in pixels and can be in the range $(1,1)$ to $(M,N)$. Normally, the meteor image is spread across multiple pixels and the meteor position can be determined by centroiding with sub-pixel accuracy. 

The timeline of the exposure is displayed in Fig. \ref{f_rs}. The time increases from left to right. The numbers on vertical axis represent the row numbers. The orange color indicates the time intervals when the exposure proceeds, i.e. all pixels of the given row collect the light. The gray color indicates the readout time when the data are read by the camera electronics from the row. The exposure cannot proceed during that time. The white color indicates idle time.

We can define the following time intervals:
\begin{itemize}
\item frame-to-frame time $(T)$ – the time interval between the start of exposures of two subsequent frames, $T = 1/F$, also known as frame time or frame period 
\item readout time $(R)$ – the time needed to read out the data from a row
\item exposure time $(E)$ – time of the exposure of one video frame, $E + R \leq T$
\item time delay of a line $(\tau)$ – time interval between the start of exposures of each two subsequent rows of the sensor
\item rolling period $(P)$ –  time required to start the exposures of all rows of the frame and be ready to start the next frame. It is the time interval between the start of the exposures of the first and the last row plus one more delay time, after which the exposure of the first row could start again, if needed. $P = \tau N, P \leq T$
\end{itemize}

\section{Correction of rolling shutter}\label{cor}
The first step in computing the meteor trajectory and velocity is determining the meteor celestial coordinates (right ascension and declination or azimuth and elevation) at each camera as a function of time. The coordinate determination (astrometry) is not discussed here. For assigning the time to each measurement, let us assume that we know the time of the start of the exposure of each video frame. Subsequent frames are separated by time $T$. In case of global shutter, the same is valid for subsequent meteor measurements. The assigned time will be the start of the exposure $+ E/2$. 

However, the movement of the meteor across the sensor with a rolling shutter causes the measured positions of the meteor in two consecutive video frames to not be separated by a time interval $T$. The time between them depends also on $\tau$ and $\Delta y$, where $\Delta y$ is the difference of meteor positions in the two frames in the $y$-coordinate.

If the exposure of the $i$-th frame starts at a time $s_i$, and the vertical position of the meteor in this frame is $y_i$, the meteor assigned time will be
\begin{equation}
t_i = s_i +  (y_i - 1) \tau + E/2
\end{equation}
In practice, we are interested in relative time for the $i$-th frame in respect to the first frame, where the meteor was visible. If the time of the first detection is $t_0$, the corresponding frame number is $f_0$, and the meteor coordinate at that frame was $y_0$, the meteor time in the $i$-th frame, $f_i$, is
\begin{equation}
t_i = t_0  + (f_i - f_0) T + (y_i - y_0) \tau
\label{e_ti}
\end{equation}
For global shutter, $\tau = 0$. For rolling shutter, we need to know $\tau$ which can be computed from the rolling period, $P$, as $\tau = P/N$.

\section{Kukić et al. (2018) correction of rolling shutter}\label{kuk}
The published procedure of rolling shutter temporal correction in \cite{kuk18} has one simplification and two mistakes. The simplification is in their equation $(10)$, which reads in our notation: $\tau = 1/(FN)$. They simply assume that the rolling period, $P$, is equal to the frame-to-frame time, $T = 1/F$.  That is possible but not generally valid. $P$ can be much shorter than $T$.

This simplified correction corresponds to the correction used by the GMN and their Raspberry Pi Meteor Station (RMS) software to analyze meteors. According to github.com\footnote{https://github.com/CroatianMeteorNetwork/RMS/blob/master/RMS/Routines/RollingShutterCorrection.py}, rolling shutter correction is implemented in the RollingShutterCorrection.py routine and does not include the value of the sensor rolling period. They define the relative offset in fractional frames $\Delta f = y/N$, where $y$ is the $y$-position of the meteor centroid, and the final corrected non-integer frame number is $f' = f + \Delta f$.  They then divide this fractional frame number by FPS to get the time for every frame. If the meteor time at zero frame was $t_0$, the time of the meteor at the $i$-th frame is
\renewcommand{\theequation}{2a}
\begin{equation}
t_i = t_0  + \frac{(f_i - f_0)}{F} + \frac{(y_i - y_0)}{FN}
\label{e_tia}
\end{equation}

\renewcommand{\theequation}{\arabic{equation}}
\setcounter{equation}{2} 
If we compare this equation with eq. \ref{e_ti}, we see that they are identical only if $\tau = 1/(FN)$.

The first mistake in \cite{kuk18} is above their equation $(12)$ in the text description of the fractional exposure. Their definition is "the ratio of exposure time to the frame-to-frame time divided by the frame rate FPS", which is not correct. Furthermore, $f$ is not named anywhere. However, equation $(12)$, i.e., the mathematical formula for $f$, is already correct. The second mistake in \cite{kuk18} is in equation $(13)$ and meaning of $t_f$. The authors mixed two different interpretations of $t_f$. First, $t_f$ is defined as the time bias of the centroid of meteor, but in the equation $(13)$, it is an exposure-related quantity. Equation $(13)$ implies that $t_f=f/FPS$, but $f/FPS$ is the definition of exposure time according to equation $(12)$, which would mean that the variable $t_f$ is equal to exposure time, which cannot be true.

\section{The influence of rolling shutter on computation of meteor velocities}\label{inf}
Rolling shutter has no influence on meteors moving along a row, with constant $y$-coordinate. The highest influence is on meteors moving vertically. Let us assume that meteor position changes by $\Delta y$ pixels between two consecutive frames. That motion corresponds to a physical distance $\Delta L$ in the atmosphere, $\Delta L = k \Delta y$, where the factor $k$ depends on the geometry of the camera and the meteor but does not depend on the properties of the camera shutter. The instantaneous meteor velocity is $v = \Delta L/\Delta t$, where the time interval between the two measurements, $\Delta t$, depends on the shutter properties. In case of global shutter, it will be simply $T$. The computed velocity assuming global shutter will be 
\begin{equation}
v_0 = \frac{k\Delta y}{T}
\end{equation}
If there is a rolling shutter with rolling period $P$ and we determine $\Delta t$ from eq. (\ref{e_ti}), the velocity will be
\begin{equation}
v_P = \cfrac{k\Delta y}{T+\frac{\Delta y}{N}P}
\end{equation}
The relative difference is
\begin{equation}
\delta v = \frac{v_0 - v_P}{v_P} = \frac{\Delta y}{N}\frac{P}{T}
\label{e_dv}
\end{equation}
As expected, the difference increases with $\Delta y$ and with $P$. Of course, if incorrect value of $P$ is used, the resulting velocity is also incorrect. While $\Delta y$ can be measured and $N$ and $T$ are well known, the value of $P$ is usually not provided by the camera manufacturer. We are interested to know how precisely must be $P$ determined not to introduce additional errors or uncertainties in meteor velocity measurements.

As we can see from eq. (\ref{e_dv}), the rolling shutter influence is not proportional just to $\Delta y$ but to the ratio   $\Delta y/N$, that means to the fraction of the vertical field of view traveled by the meteor between two frames. The smaller field of view, the more crucial is the rolling shutter correction. It is, of course, more important for meteors with large angular speed. As a reasonable example, we can consider a meteor moving by 1.5$^\circ$ within the frame-to-frame time 0.04\,s (camera with 25 frames per second). It corresponds to a meteor with velocity 65\,km/s moving perpendicularly to the line of sight at a distance of 100\,km.

The precision of velocity measurement within the European Fireball Network is certainly better than 1$\%$. We therefore request the uncertainty introduced by the rolling shutter to be lower than 10$^{-3}$. The uncertainty of the rolling period must be
\begin{equation}
\delta P < \cfrac{10^{-3}}{\frac{\Delta y}{N}}T
\label{e_dP}
\end{equation}
For a camera with vertical field of view of 30$^\circ$, we have $\Delta y/N $= 0.05. With $T = $40\,ms, we get $\delta P < $0.8\,ms. Although this was an extreme example, we certainly need to know the rolling period with the precision of at least 1\,ms.

\section{Actual example}\label{act}
\begin{figure}[h]
\centering
\includegraphics[width=0.99\textwidth]{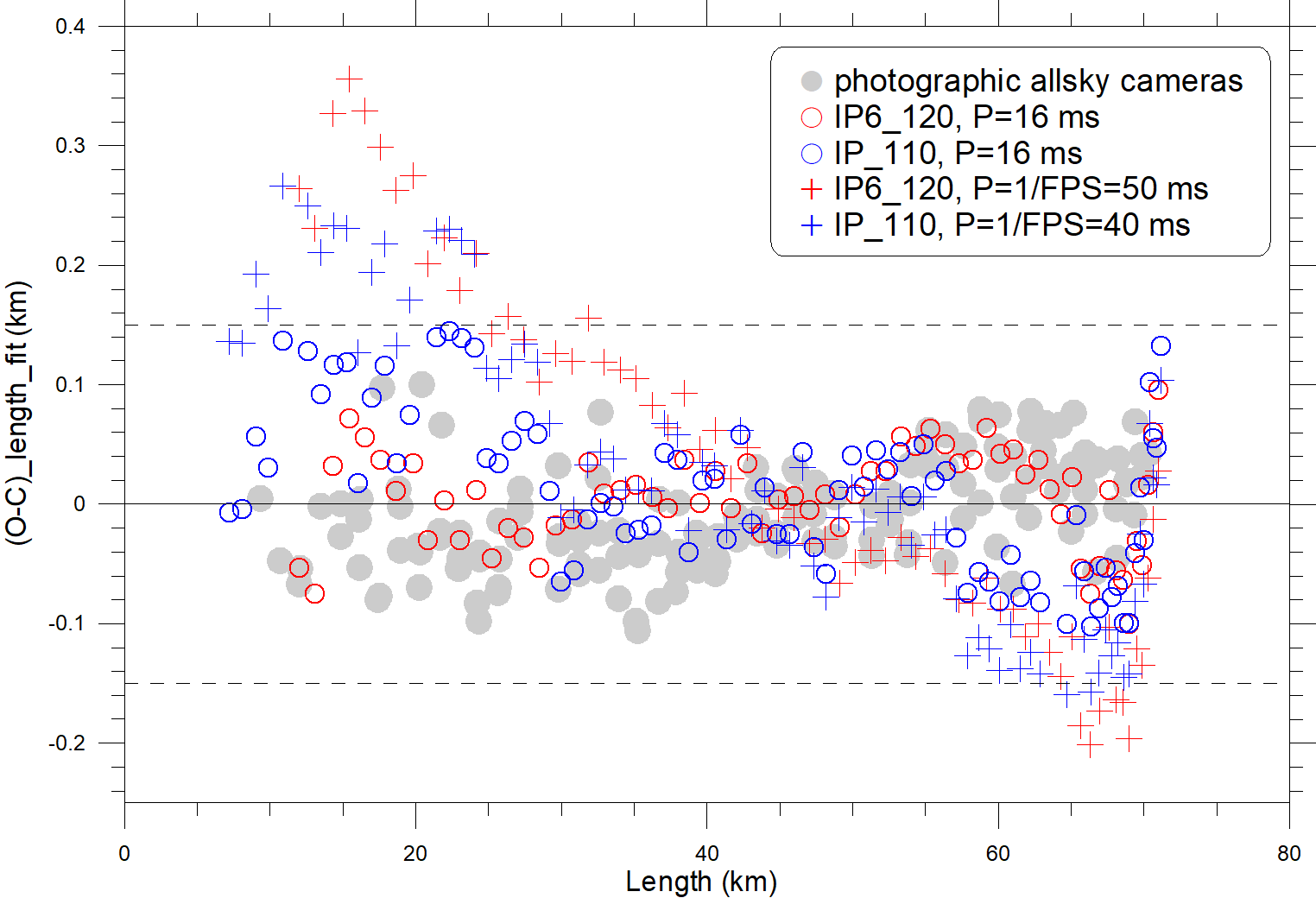}
\caption{Comparison of two rolling shutter corrections applied on two video records of the EN021124\_214006 fireball observed by the CFN in November 2024. The dashed lines show the 3-sigma interval of $\pm$\,0.15\,km.}
\label{f_O-C}
\end{figure}

Figure \ref{f_O-C} shows a comparison of corrections according to \cite{kuk18} and according to this work. We used the EN021124\_214006 fireball observed by CFN photographic and video cameras in November 2024. The initial velocity of the fireball was 22\,km/s, the terminal velocity was 5\,km/s,  the duration of the fireball was 3.4\,s. Figure \ref{f_O-C} shows the deviations of individual measurements from the physical four-parameter fit to the length-data based on the integrals of single-body differential equations of meteoroid deceleration and ablation \citep{pec83}. Four photographic cameras are shown as gray circles, two IP cameras with time corrected according to the CFN procedure (equation \ref{e_ti}) as red and blue empty circles and the same two IP cameras corrected according to the \cite{kuk18} procedure (equation \ref{e_tia}) as red and blue crosses. The red color indicates the IP6\_120 camera with 20\,FPS and a vertical field of view (FOV) of 32 degrees, while the blue color indicates the IP\_110 camera with 25\,FPS and a vertical FOV of 48 degrees. The numbers 120 and 110 are the designations of the CFN Ondřejov and Polom stations respectively. It can be seen that corrections including the rolling period give dynamics consistent with that determined from photographic records. If the incorrect rolling shutter correction is applied, the residua show systematic trend, being positive at the beginning and negative at the end. The usage of such data without photographic records would lead to a velocity lower that it actually was.

\section{Determination of the rolling period}\label{det}
\begin{figure}[h]
\centering
\includegraphics[width=0.99\textwidth]{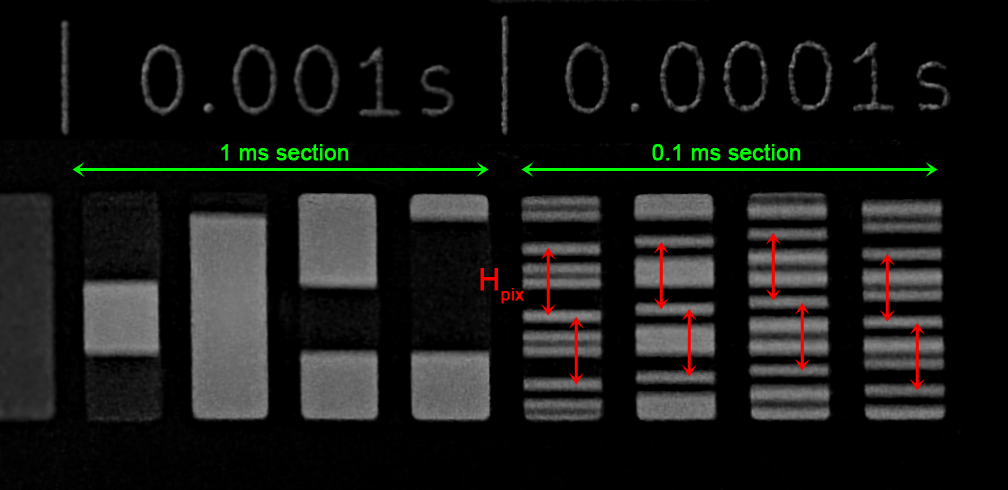}
\caption{NEXTA device diodes showing repeating patterns in 0.1\,ms section. The height of the repeating patterns $H_{pix}$ (red arrows) is inversely proportional to the readout time.}
\label{f_NEX}
\end{figure}

We have tried four methods to determine the rolling period values of our video cameras. Two of them require continuous rotation of the IP camera or the captured image (light spot created by a laser), one is based on stroboscope illumination and one on time-synchronized blinking diodes. Errors in determining $P$ values ranged from tenth to tens of milliseconds.

The first method for determining $P$ that we tried is described on the website guthspot.se\footnote{https://www.guthspot.se/video/deshaker.htm}. The method uses five measured positions of a vertical line during rest and regular camera rotation. We tried several camera placements and two rotation motors, but due to inaccuracies in rotation, assembly vibration, and lens distortion at the edge of the FOV, the $P$ value for the HFW4431 IP camera was 19 $\pm$ 44\,ms, which makes this method unreliable.

The second method we used is based on measuring the distance between the positions of the moving light spot in two consecutive video frames. The spot first moves from top to bottom and then from bottom to top, i.e., either in the direction of or against the direction of reading the sensor rows. The speed of motion is arbitrary but must be the same in both cases. If the speed is $w$ rows per second, the spot will move by $n_0 = wT$ rows within the frame-to-frame interval. However, due to the rolling shutter, the measured distance between the two positions will be longer when moving down $(n_2 > n_0)$ and shorter when moving up $(n_1 < n_0)$. The following relations will be valid:
\begin{equation}
\begin{split}
n_2 &= w (T + n_2 \tau) \\
n_1 &= w (T - n_1 \tau)
\end{split}
\label{e_n12}
\end{equation}
Eliminating $w$, we will arrive at
\begin{equation}
P = \tau N = \frac{(n_2 - n_1)}{2n_2n_1}TN
\end{equation}
In this setup, the camera is fixed and the laser is mounted on a rotary motor that rotates in one direction or the other. We set the IP camera to the shortest possible exposure time so that the light spot of the laser trace is circularly symmetric, and repeated the measurements for different laser rotation speeds. We obtained $P$ value for the HFW4431 IP camera of 16.6 $\pm$ 1.8\,ms. This method provides reliable results but our setup was probably affected by the assembly vibrations and changing distance of the laser light spot from the camera (wall projection), which increased the error of the $P$ value.

The third method of measuring $P$ is based on controlled flashing of a light source with a short flash duration (1\,ms and shorter) and a known delay between the flashes. We were inspired by the stroboscope illumination method described in \cite{bra09}. We recorded a ceiling illuminated by light-emitting diodes. The change in the position of the edge between the illuminated and unilluminated parts of the FOV is then measured on the recording and its vertical shift is a function of the rolling period. For the HFW4431 IP camera, we obtained two different $P$ values for two different flash duration and delay settings, namely 15.7 and 14.2\,ms. However, both values were determined with an error of less than 0.1\,ms, which means that not all effects of flash duration were taken into account and the determined $P$ value is not entirely correct.

The only method accurate enough to determine $P$ with an error of tenths of milliseconds was the last one we tried. It involves the use of a New EXposure Timing Analyser (NEXTA) designed for astronomical observations timed with submillisecond accuracy \citep{kam23}. The NEXTA device is a strip of blinking diodes that are synchronized using the Global Navigation Satellite System and are capable of displaying the time with an accuracy of 0.1\,ms. The NEXTA device we used was on long-term loan from one of the authors of this work (J.M.) for testing and measurement purposes. Recording these diodes using a rolling shutter camera with exposure time of single video frame shorter than one thousandth of a second shows repeating patterns of a certain height in pixels, $H_{pix}$ (Figure \ref{f_NEX}), which is a function of the rolling period. It is
\begin{equation}
P = \frac{10\,t_{nexta}N}{H_{pix}},
\label{e_P}
\end{equation}
where $t_{nexta}$ is a section of diodes on the NEXTA device, where the $H_{pix}$ is measured (0.1\,ms section in our case, so $10\,t_{nexta}=$ 1\,ms). We always measured several dozen values of $H_{pix}$, (pairs of positions defining the $H_{pix}$,) and then determine the mean value and its variance. The 0.1\,ms section of diodes on the NEXTA devise has an area of approximately 1\,cm$^2$. It is not possible to record such a small area with cameras with wide-angle lenses focused on infinity at sufficient resolution to calculate the $P$ value with sufficient accuracy. Therefore, we placed a 50\,mm photographic lens between the video camera and the NEXTA device. It was thus possible to capture video footage of the diodes using cameras with a field of view between 50 and 110 degrees for the same setup.

We obtained $P$ value for the HFW4431 IP camera of 16.27 $\pm$ 0.06\,ms. We measured the $P$ values for all IP cameras used in CFN and two cameras used by Allsky7 network in this way. Their values are shown in Table \ref{t_rol}.  The values are often around 16 or 32\,ms, which is probably connected with the ability of the sensor to take either 60 or 30 frames per second. But other values are present as well and measurements are needed for each individual camera and each configuration. We can see that changing the camera resolution may or may not change the rolling period. On the other hand, rolling period remains the same if exposure time is changed.

\begin{table}[h]
\caption{Rolling periods determined by the NEXTA device.}
\label{t_rol}%
\begin{tabular}{@{}lcccc@{}}
\toprule
 & resolution  & FPS & 1/FPS & P\\
 & (pix) & & (ms) & (ms)\\ 
\midrule
Dahua IP (EN) &&&&\\
HFW4421      & 2688$\times $1520 & 20 & 50 & 16.20 $\pm$ 0.10 \\
HFW4431      & 2688$\times $1520 & 25 & 40 & 16.27 $\pm$ 0.06 \\
HFW3841      & 3840$\times $2160 & 25 & 40 & 31.99 $\pm$ 0.18 \\
                    & 2688$\times $1520 & 25 & 40 & 24.66 $\pm$ 0.28 \\
HFW3841-S2 & 3840$\times $2160 & 25 & 40 & 31.99 $\pm$ 0.22 \\
                    & 2688$\times $1520 & 25 & 40 & 32.21 $\pm$ 0.19 \\
HFW5442      & 2688$\times $1520 & 25 & 40 & 14.01 $\pm$ 0.09 \\
HFW3441      & 2688$\times $1520 & 25 & 40 & 32.63 $\pm$ 0.22 \\
HFW3441-S2 & 2688$\times $1520 & 25 & 40 & 38.66 $\pm$ 0.29 \\
HFW3449-IL  & 2688$\times $1520 & 25 & 40 & 15.87 $\pm$ 0.09 \\
HFW3541      & 2688$\times $1520 & 25 & 40 & 32.98 $\pm$ 0.35 \\
\midrule
Sony (Allsky7) &&&&\\
IMX291        & 2048$\times $1536 & 25 & 40 & 32.05 $\pm$ 0.22 \\
                   & 1920$\times $1080 & 25 & 40 & 32.20 $\pm$ 0.24 \\
IMX307        & 2048$\times $1536 & 25 & 40 & 31.95 $\pm$ 0.30 \\
                   & 1920$\times $1080 & 25 & 40 & 32.13 $\pm$ 0.32 \\
\botrule
\end{tabular}
\end{table}

\section{Conclusions}\label{con}
Accurate determination of meteor dynamics from video cameras that use rolling shutter sensors requires the use of corrections that require the knowledge of the rolling period. We demonstrated that ignoring the correction or using inappropriate methods or values produces unnecessary errors. It makes no sense to speculate for which geometry of meteor and camera such a correction is needed or how large the error of velocity determination is if this correction is not applied. All meteor observers should want to determine meteor parameters with the highest possible accuracy and therefore not use simplified formulas as in the work of \cite{kuk18}. We provided the correct formula (eq. \ref{e_ti}), and using it is essential for determining accurate results for meteors captured by this type of video camera.

Unfortunately, the values of rolling periods are not part of the manufacturer's camera specification in most cases. We measured the values just for several cameras. For other cameras rolling periods must be determined individually. We recommend using the NEXTA device described in \cite{kam23}. 

\backmatter

\bmhead{Acknowledgements}
We would like to thank Mike Hankey and Sirko Molau for lending us the Allsky7 camera modules. This work was supported by the institutional project RVO: 67985815. 

\bmhead{Author Contributions}
L.S. wrote the original draft; reviewed and edited the manuscript, measured $P$ values. J.B. proposed one method for measuring $P$, reviewed, edited and extended the original draft. P.S. calculated EN fireballs and empirical $P$ values. J.F. and J.M. helped measure $P$ values. All authors commented on previous versions of the manuscript. All authors read and approved the final manuscript.

\bmhead{Data Availability}
Laboratory data used and analysed during this work are available from the corresponding author upon reasonable request.


\bmhead{Competing interests}
The authors declare no competing interests.






\bibliography{RS_Shrbeny}

\end{document}